# Self-Consistent Particle Acceleration in Active Galactic Nuclei

A. Mastichiadis and J. G. Kirk

Max-Planck-Institut für Kernphysik, Postfach 10 39 80, 69029 Heidelberg, Germany

**Abstract.** Adopting the hypothesis that the nonthermal emission of Active Galactic Nuclei (AGN) is primarily due to the acceleration of protons, we construct a simple model in which the interplay of acceleration and losses can be studied together with the formation of the emitted spectrum. The acceleration process is assumed to be of the first order Fermi type, and the proton distribution as well as the injected electrons and photons in the central region of the AGN are described by spatially averaged kinetic equations. The various relevant processes which dominate the three species are incorporated into the equations. The technique used to solve these is presented and several tests of the numerical implementation are presented. We also present results of a sample time-dependent AGN model in which photons appear suddenly as a result of a feedback instability and the system evolves to a steady state, in which the acceleration process is saturated self-consistently by the photons it produces. This example combines an X-ray power law index of about $-1.7$, together with a break at an energy between 50 and 500 keV.

**Key words:** Acceleration of particles; Galaxies: active, nuclei, Seyfert; Gamma rays: theory; X-rays: galaxies

## 1. Introduction

Active Galactic Nuclei are thought to be powered by accretion of matter onto a central massive black hole (Rees 1984). However, the way in which gravitational energy is converted into electromagnetic radiation remains largely unclear. Studies of the X-ray and $\gamma$-ray properties of AGNs can be particularly illuminating as variability (especially in the well-measured X-ray regime) implies that this radiation is generated within a few Schwarzschild radii of the black hole (Done & Fabian 1989).

Over the last decade, considerable effort has been expended on understanding the X-ray spectrum of AGNs. The high values which X-ray observations imply for the photon compactness $\ell_\gamma = L_\gamma \sigma_T / 4\pi R m_e c^3$ (where $L_\gamma$ is the luminosity, $R$ the radius of the source and $\sigma_T$ the Thomson cross-section) mean that any $\gamma$-ray of energy $> 1$ MeV, if indeed present, would be readily absorbed by the softer photons. The resulting electron-positron pairs would quickly lose their energy, producing more energetic photons and an intense electromagnetic cascades would ensue (for a review see Svensson 1989 and references therein). Such a cascade could be expected to alter drastically the primary photon spectrum, and it was hoped that this property might explain the canonical X-ray power law spectral index of $-1.7$, as observed by both the HEAO-1 (Rothschild et al. 1983) and the GINGA (Turner & Pounds 1989) X-ray observatories. However, these models do not include a discussion of particle acceleration, but instead describe the required injection of energetic electrons and photons by a number of free parameters. It turns out that only a rather limited region in this parameter space gives acceptable results (Lightman & Zdziarski 1987) – a conclusion which is, however, relaxed by the introduction of models including reflection (see Pounds et al. 1989 and Zdziarski et al. 1990). Another shortcoming of pair cascade models is that although variability is one of the basic characteristics of AGNs most of them assume steady state conditions. Without a model of acceleration, it is possible to investigate only rather artificial variations of the injection paratmeters (Done & Fabian 1989, Coppi 1992).

Approaching the problem from a different angle, a class of models proposing the presence of a strong stationary shock in the central part of an AGN has been developed (Protheroe & Kazanas 1983, Kazanas & Ellison 1986). Such a shock can, in principle, accelerate protons to very high energies, thereby converting part of the gravitational energy released by the plasma accreting into non-thermal particles. The presence of high energy protons in AGNs has many interesting consequences, which have been investigated in a number of papers: relativistic protons will eventually lose part of their energy in inelastic collisions with ambient protons (Protheroe & Kazanas 1983, Kazanas & Ellison 1986) or ambient photons (Sikora et al. 1987). In each case neutrons and neutrinos, as well as high energy electrons/positrons and $\gamma$-rays will be produced. Unlike protons, which are magnetically confined inside the acceleration region, the neutrons escape, carrying off a fraction of the initial luminosity (Kirk & Mastichiadis 1989, Sikora et al. 1989, Atoyan 1992a). Various observable or potentially observable effects of these neutrons have been predicted, such as the production of Boron in spallation reactions (Kirk & Mastichiadis 1989), the emission of Very High Energy $\gamma$-rays (Mastichiadis & Protheroe 1990, Atoyan 1992b), the acceleration of winds in broad absorption line QSOs (Begelman et al. 1991) and the production of cosmic rays around the 'knee' (Protheroe & Szabo 1992). Also, the flat radio spectra of radio loud AGNs have been attributed indirectly to escaping

*Send offprint requests to*: A. Mastichiadis

neutrons (i.e., after their decay into protons and electrons – Giovanoni & Kazanas 1990). Finally, the flux of escaping neutrons has been used to calculate the contribution of AGNs to the diffuse $\gamma$-ray background (Johnson et al. 1994). Neutrinos, on the other hand, may be produced at a rate which could be detected with an experiment such as DUMAND (Stecker et al. 1991, Biermann 1992, Sikora & Begelman 1992, Szabo & Protheroe 1992, 1994). In these models the relativistic protons also inject a population of electrons and positrons, which cool by synchrotron and/or inverse Compton radiation and initiate intense pair cascades. However, the resulting nonthermal radiation (which is, at least partly, responsible for the saturation of the acceleration) was specified a priori and not calculated self-consistently. In this sense, the hadronic models include a model of particle acceleration and injection but lack a self-consistent treatment of the electromagnetic cascades.

A first attempt to create a synthesis of these models was made by Stern et al. (1991), Stern & Svensson (1991) and Stern et al. (1992). These authors followed the electromagnetic cascades resulting from the injection of electrons by relativistic protons and included the feedback of the photons on the relativistic protons. They found that this system showed limit cycles much like a predator-prey system. However, the Monte Carlo approach which they use complicates the interpretation of the results in terms of a specific feedback mechanism. Such a feedback effect was found analytically by Kirk & Mastichiadis (1992 – henceforth "KM92") using the kinetic equation approach. They showed that relativistic protons become unstable to a combination of proton-photon pair production and electron synchrotron radiation once their number density and energy exceeds a certain critical value. This feedback leads eventually to rapid energy losses for the relativistic protons. However, the analysis of KM92 employs a stationary proton distribution and uses a linearised set of equations which exclude electromagnetic cascades, making it impossible to follow the evolution of the system into the saturation phase.

Motivated by this shortcoming we present here a model capable of describing time-dependent effects in the production of the nonthermal spectra of AGNs, albeit under a set of highly simplifying assumptions. Our method is to describe the three basic components of the central region of an AGN – protons, electrons and photons – by a system of three spatially averaged kinetic equations. The aim is to incorporate in an approximate manner all the important processes acting on these constituents and to use numerical methods to integrate the system forwards in time in the manner described by Fabian et al. (1986) and Coppi (1992) for photons and electrons. A time-dependent method is also essential if one is to account for the highly nonlinear coupling of the acceleration process with losses caused by the associated photons. We propose to use this technique to gain a better understanding of the origin and properties of variability in AGNs as well as the way in which their photon spectra are formed. Our method should also provide better information about quantities such as the expected luminosity in high energy neutrinos. In the present paper, however, we restrict ourselves to a detailed description of the method together with a discussion of its strengths and weaknesses. We do not attempt a systematic investigation of the parameter regime appropriate to AGNs, but present a sample set of results obtained using the full code. Preliminary results of this work have been presented by Mastichiadis & Kirk (1992).

The contents of the paper are organised as follows: in Sect. 2 we describe the way in which we model the acceleration process. Only protons are assumed to undergo acceleration – it being tacitly assumed that the relativistic electrons are dominated by rapid loss processes. We choose a model in which a first-order partial differential equation is used to describe the first-order Fermi process and a 'loss' or 'escape' probability is introduced to account for the possibility that protons might leave the emission region, for example by accretion into the black hole. In such a model, acceleration occurs homogeneously throughout the emission region, as might be expected, for example, if a converging accretion flow provides the acceleration (Schneider & Bogdan 1989).

Apart from acceleration, the microscopic processes of importance in the system of kinetic equations are relatively well understood. For the protons these include proton-proton and proton-photon collisions (Mannheim & Biermann 1989, Begelman et al. 1990) whereas the electrons and photons (in addition to the source terms provided by the proton related processes) experience synchrotron radiation, Compton scattering, photon-photon pair production, electron-positron annihilation and Compton downscattering on cooled electrons (Coppi & Blandford 1990). These processes and the approximations we employ to describe them are discussed in Sect. 3.

The method used to convert the integro-differential kinetic equations into a system of ordinary differential equations suitable for integration by standard numerical methods is presented in Sect. 4, and this is followed by a series of tests which check the behaviour of our approximation schemes in circumstances in which either analytic solutions are available e.g., when only synchrotron cooling or inverse Compton cooling are present, or in which there exist calculations in the literature with which to compare, e.g., the spectra of stationary electromagnetic cascades in the Thomson regime (Lightman & Zdziarski 1987).

As an example of the application of the full code, Sect. 5 presents time-dependent results for one particular set of parameters which are appropriate for an AGN. One of the features of this run is the development of the pair production-synchrotron instability (KM92) once the marginal stability threshold is crossed. The X-ray spectrum of power-law index $\alpha = -5/3$ which is predicted in the linear phase of the instability is seen to persevere well into the nonlinear phase and a stationary state is achieved in which the photons produced by accelerated protons are responsible for saturating the acceleration process. To conclude, we briefly summarise the main advantages and limitations inherent in our approach in Sect. 6.

## 2. The Particle Acceleration Model

According to the theory of particle acceleration by the first-order Fermi mechanism (see, for example, Kirk et al. 1994) stochastic encounters of particles with so-called 'scattering centres' occur within an 'acceleration region', such that each interaction results in a small increase in the particle's energy. At the same time, a particle has a finite probability of escaping from the acceleration region, and the combination of these two effects leads to a distribution of accelerated particles which, under certain circumstances, is of the power-law type. We shall adopt this theory here, and make the simplest possible assumptions concerning the rate at which energy is gained from the scattering centres as well as the rate at which parti-cles escape the region containing these centres. The aim is to describe the particle distribution using a kinetic equation in which not just acceleration, but also losses suffered as a result of interactions with ambient photons can be included. However, one important question about the nature of an AGN, which we cannot answer a priori is that of how large the acceleration region is compared to the size of the 'source region', i.e., that region which contains the relativistic particles responsible for the observed nonthermal emission. Two possibilities represent opposite extremes:

1. acceleration occurs throughout the whole of the source region, and
2. acceleration takes place in only very small parts of the source.

An example of case 1 is the acceleration of particles by a smooth, unshocked accretion flow (Payne & Blandford 1981, Cowsik & Lee 1982, Webb & Bogdan 1987, Schneider & Bogdan 1989). Particles may escape from the acceleration region in this case (perhaps by being accreted into the black hole) but if they do, they cease to emit observable radiation and, by definition, have left the source. An example of case 2 is the acceleration of particles at a shock front in the accretion flow (Protheroe & Kazanas 1983, Sikora et al. 1987). Provided the mean free path of the particle is small compared to the source, only those particles in the immediate vicinity of the shock undergo acceleration. Once particles have been swept out of this region, they are highly unlikely to return to it, but may still be energetic enough to produce observable radiation whilst cooling on the ambient photons, and so cannot be considered to have left the source. In this paper we deal exclusively with case 1, in which scattering centres are distributed throughout the source region.

Denoting by $\hat{n}_p(p)dp$ the differential number density of protons of momentum $p$ in the interval $dp$, the kinetic equation for protons within the source region can be written (e.g., Schlickeiser 1984, Kirk et al. 1994)

$$\frac{\partial \hat{n}_p(p,t)}{\partial t} + \frac{\partial}{\partial p}\left[\frac{p}{\hat{t}_{acc}}\hat{n}_p(p,t)\right] + \frac{\hat{n}_p(p,t)}{\hat{t}_{esc}}$$
$$= Q_{inj}\delta(p - p_{inj}) + \hat{\mathcal{L}}^p(\hat{n}_p, p, t) \qquad (1)$$

where $Q_{inj}$ is the number of particles injected into the acceleration process per second per unit volume with momentum $p_{inj}$ and $\hat{\mathcal{L}}^p$ (which can be a differential and/or integral operator acting on $\hat{n}_p$) denotes the losses suffered by energetic protons. The second term in Eq. (1) provides a continuous energy input by the first-order Fermi process into those protons which remain in the acceleration region, whereas the third allows for escape, the average residence time being $\hat{t}_{esc}$.

If we ignore for a moment the losses, and take the simple case of constant $Q_{inj}$, $\hat{t}_{acc}$, and $\hat{t}_{esc}$, the solution to Eq. (1) which satisfies the boundary condition $\hat{n}_p(p, 0) = 0$ is (Axford 1981)

$$\hat{n}_p(p,t) = \frac{\hat{t}_{acc}Q_{inj}}{p_{inj}}\left(\frac{p}{p_{inj}}\right)^{-1-\hat{t}_{acc}/\hat{t}_{esc}} \times$$
$$\left[H(p - p_{inj}) - H(p - p_{inj}e^{t/\hat{t}_{acc}})\right] \quad , \qquad (2)$$

where $H(x)$ is the Heaviside function, equal to zero for $x < 0$ and unity for $x > 0$. This solution is just a power-law extending from $p_{inj}$ up to a cut-off at $p_{max}(t) = p_{inj}e^{t/\hat{t}_{acc}}$ which increases with time. Below the cut-off i.e., for $p < p_{max}(t)$ the solution is independent of time. The power-law index is determined by the relative strengths of the acceleration term and escape term.

In the following we will usually assume $\hat{t}_{acc} = \hat{t}_{esc}$, in which case $\hat{n}_p \propto p^{-2}$, such as is expected, for example, of particles accelerated at a strong shock front in a gas of adiabatic index 5/3. Interesting conclusions can be drawn in this special case about the energy given to and extracted from the protons by building the moment of Eq. (1) with the kinetic energy of a proton: $(\gamma - 1)m_p c^2$, where $\gamma$ is the Lorentz factor ($\gamma = \sqrt{1 + p^2/(m_p c)^2}$). Denoting the total energy (minus the rest mass) contained in accelerated protons by $E$, we find after integrating by parts:

$$\frac{dE}{dt} = VQ_{inj}m_p c^2(\gamma_{inj} - 1) - L_{loss}$$
$$+ Vm_p c^2 \int_0^\infty dp \frac{(\gamma - 1)\hat{n}_p(p,t)}{\gamma \hat{t}_{acc}} \quad , \qquad (3)$$

where $\gamma_{inj} = \sqrt{1 + p_{inj}^2/(m_p c)^2}$, $L_{loss}$ is the rate at which energy is lost by protons (to the processes of pair production and pion production) and $V$ is the source volume. The loss processes discussed in the following section operate effectively only on relativistic protons, so we can expect that if a steady state is set up in which $E =$ constant, the spectrum will be given by the loss-free solution Eq. (2) all the way from the injection momentum up into the relativistic regime. However, the integral in the third term on the right-hand side of Eq. (3) is dominated by the nonrelativistic and transrelativistic regimes, provided the particle density does not diverge at large $p$. Consequently, we make only a small error by using the loss-free distribution in this integral. Provided $p_{inj} \ll m_p c$, we find in the steady state:

$$L_{loss} = Vcp_{inj}Q_{inj}. \qquad (4)$$

According to this equation, the rate at which protons put energy into pair production and pion production during the acceleration process (which is in this model the entire nonthermal luminosity of the AGN) is determined in the steady state solely by the rate at which they are injected into the acceleration process at low momentum. In particular, it is independent of quantities connected with the actual loss process itself, such as the background photon or matter density and the maximum Lorentz factor to which particles can be accelerated, even though these may depend nonlinearly on the density itself. In connection with Eq. (4), we note that the rate $L_{inj}$ at which energy is injected at momentum $p_{inj}$ is small compared to the nonthermal luminosity: $L_{inj}/L_{loss} = p_{inj}/2m_p c \ll 1$, so the nonthermal emission stems not from the unknown injection process, but from the first-order Fermi mechanism we are modelling.

In this study of AGNs, we will be concerned only with relativistic protons, since it is in this regime that the acceleration and loss processes can compete with each other. It is then more convenient to write the kinetic equation (1) in terms of the Lorentz factor $\gamma$ of the protons, using a normalisation in which time is measured in units of the light crossing time of the source (of size $R$) and the particle density refers to the number contained in a volume element of size $\sigma_T R$ (where $\sigma_T$ is the Thomson cross section). Accordingly, we define the di-

mensionless density $n_p(\gamma, t)$ by

$$n_p(\gamma, t)d\gamma = \sigma_T R \hat{n}_p(p, t) dp \qquad (5)$$

and find, in the relativistic regime:

$$\frac{\partial n_p(\gamma, t)}{\partial t} + \frac{\partial}{\partial \gamma}\left[\frac{\gamma}{t_{acc}} n_p(\gamma, t)\right] + \frac{n_p(\gamma, t)}{t_{esc}} = \mathcal{L}^p(n_p, \gamma, t). \qquad (6)$$

where $t$ is now dimensionless, $t_{acc} = c\hat{t}_{acc}/R$ and $t_{esc} = c\hat{t}_{esc}/R$. The nonrelativistic part of the acceleration process can be avoided by using the loss-free solution as a boundary condition

$$n_p(\gamma_0, t) = n_0/\gamma_0^{1+t_{acc}/t_{esc}}, \qquad (7)$$

with

$$n_0 = \sigma_T R \hat{t}_{acc} Q_{inj}(p_{inj}/m_p c)^{t_{acc}/t_{esc}}, \qquad (8)$$

where $\gamma_0$ is the lowest value of the Lorentz factor to be considered. For $t_{acc} = t_{esc}$, the AGN luminosity in a steady state can be expressed in terms of the compactness

$$\ell_{tot} = \frac{L_{loss} \sigma_T}{4\pi R m_e c^3}. \qquad (9)$$

Setting the source volume $V = \frac{4}{3}\pi R^3$ in Eq. (4) leads to

$$\ell_{tot} = \frac{n_0}{3 t_{acc}}\left(\frac{m_p}{m_e}\right). \qquad (10)$$

Using the same normalisation, we complement Eq. (6) by writing for the relativistic electrons and positrons:

$$\frac{\partial n_e(\gamma, t)}{\partial t} = Q^e(n_e, \gamma, t) + \mathcal{L}^e(n_e, \gamma, t) \qquad (11)$$

Here $\mathcal{L}^e$ denotes the various electron loss terms while $Q^e$ are the injection terms. Note that there is no acceleration or escape term included in the electron equation, since we assume the loss terms to be much larger.

In our numerical treatment, we impose a lower limit $\gamma_{min}$ on the Lorentz factor of the relativistic electron population and assume particles which cool through this boundary join a population of cold electrons whose number density $N_e^{cool}(t)$ (i.e., the number in a volume $\sigma_T R$) is determined by the equation

$$\frac{dN_e^{cool}(t)}{dt} = Q^{e,cool}(n_e, t) + \mathcal{L}^{e,cool}(n_e, t) \qquad (12)$$

Contributions to the source term $Q^{e,cool}(n_e, t)$ arise from synchrotron and Compton cooling, while the contributions to the sink term $\mathcal{L}^{e,cool}(n_e, t)$ come mainly from electron-positron annihilation.

Finally, the spatially averaged photon equation reads:

$$\frac{\partial n_\gamma(x,t)}{\partial t} + \frac{n_\gamma(x,t)}{t_{\gamma esc}} = Q^\gamma(n_\gamma, x, t) + \mathcal{L}^\gamma(n_\gamma, x, t), \qquad (13)$$

where $x$ is the dimensionless photon frequency: $x = h\nu/(m_e c^2)$. Photons leave the source on the timescale $t_{\gamma esc}$ (measured in units of the light crossing time $t_{cross} = R/c$) and, rather than being advected into the black hole, are assumed to propagate freely after escape. The terms $\mathcal{L}^\gamma$ and $Q^\gamma$ denote the sinks and sources of photons. Because of the normalisation used, the quantity $\int dx\, x n_\gamma$ is simply related to the photon compactness (Guilbert et al. 1983) given by

$$\ell_\gamma = \frac{L_\gamma \sigma_T}{4\pi R m_e c^3}. \qquad (14)$$

For a spherical source we have

$$\ell_\gamma = \frac{1}{3 t_{\gamma esc}} \int dx\, x n_\gamma. \qquad (15)$$

## 3. The Physical Processes

We proceed now to discuss the various contributions to the kinetic equations (6), (11), (12) and (13) of the physical processes by which the components of the system interact with each other and with the magnetic field.

### 3.1. Proton-proton interactions

Inelastic proton-proton collisions act as an energy loss mechanism for relativistic protons. They also inject $\gamma$-rays, relativistic electrons, and neutrinos resulting from the decay of the produced neutral and charged pions. While one can calculate in detail the spectra of the products once the relativistic proton distribution is given (e.g., Dermer 1986, Mastichiadis & Protheroe 1990), for the present calculation it suffices to use the $\delta$-functional approximation to the differential cross-section for production of energetic pions $d\sigma(E_\pi)/dE_\pi = \sigma_{pp}^0 \sigma_T \delta(E_\pi - 0.15\,\gamma m_p c^2)$ where $\sigma_{pp}^0 \simeq 0.06$ is the proton-proton cross section (e.g., Atoyan 1992a).

Thus, to treat the proton losses we write

$$\mathcal{L}_{pp}^p(\gamma, t) = \sigma_{pp}^0 n_p^{targ} \left[n_p(\gamma', t) - n_p(\gamma, t)\right] \qquad (16)$$

where $n_p^{targ}$ is the density of target protons and we have assumed that each proton-proton collision removes protons from the energy bin $\gamma$ and $\gamma + d\gamma$ but injects them there from higher energy bins of energy $\gamma' = \gamma/(1 - k_{pp})$. Here $k_{pp}$ is the proton inelasticity and it is taken to be $k_{pp} = .45$

Since the mean energy per gamma-photon produced at the decay of the $\pi^0$-meson is $0.5 E_\pi = 0.075 \gamma m_p c^2$, the photon production spectrum, in this approximation, can be taken as

$$Q_{pp}^\gamma(x, t) = \frac{2}{0.075}\sigma_{pp}^0 n_p^{targ} n_p(\gamma_1, t) \qquad (17)$$

where $\gamma_1 = x m_e/(0.075 m_p)$. The above approximation essentially injects photons with a slope equal to the relativistic proton spectrum at energies greater than $\simeq 70$ MeV. It does not treat correctly injection at energies below this value. However, detailed calculations (e.g., Dermer 1986) have shown that the photon spectrum flattens considerably at energies below 100 MeV, so that the approximation introduced above is adequate for our purposes.

We turn next to the injected electrons (or positrons). Since the electron (positron) carries off, on average, 26% of the initial pion energy, it follows that $\langle E_e \rangle \approx 0.039\, E_p$ (Atoyan 1992a) and the corresponding injection spectrum resulting from the decay of charged pions is given by

$$Q_{pp}^e(\gamma, t) = \frac{2}{0.039}\sigma_{pp}^0 n_p^{targ} n_p(\gamma_2, t), \qquad (18)$$

where $\gamma_2 = \gamma m_e/(0.039 m_p)$.

Using similar arguments one can calculate the neutrino emissivity resulting from proton-proton collisions. Thus, we write

$$\mathcal{Q}_{pp}^\nu(E_\nu, t) = \frac{2}{0.037} \sigma_{pp}^0 n_p^{targ} n_p(\gamma_3, t) , \quad (19)$$

where $\gamma_3 = E_\nu/(0.037 m_p c^2)$.

It is easy to show using the above relations that the energy lost per unit time by protons is equal to the amount of energy per unit time given to photons, electrons and neutrinos, i.e., our approximate treatment conserves energy.

Finally, it is interesting to note that if we assume the proton losses in Eq. (16) are catastrophic, i.e., if we write

$$\mathcal{L}_{pp}^p(\gamma, t) = -\sigma_{pp}^0 n_p^{targ} [n_p(\gamma, t)] , \quad (20)$$

then instead of Eq. (16), we obtain essentially Eq. (6) and can immediately write the analytic solution:

$$n_p(\gamma, t) = n_0 \left(\frac{\gamma_{inj}}{\gamma}\right)^{1+(t_{acc}/t_{esc})+n_p^{targ}\sigma_{pp}^0 t_{acc}} \times$$
$$\left[ H(\gamma - \gamma_{inj}) - H(\gamma - \gamma_{inj} e^{t/t_{acc}}) \right] . \quad (21)$$

Comparing the above solution with Eq. (2) we see that the inclusion of proton-proton losses causes the proton distribution to become steeper, but does not affect the time evolution of the upper cut-off.

### 3.2. Proton-photon pair production

A photon of (dimensionless) energy $x$ can produce an electron/positron pair in the Coulomb field of the proton if the threshold condition

$$\gamma_p x \geq 2 \quad (22)$$

is satisfied. Assuming that the resulting electron and positron have the same Lorentz factor as the incoming proton (which is true provided the proton-photon collisions occur predominantly close to threshold), the fractional energy loss of the proton is small. In this case the losses can be considered a continuous process and can be written as

$$\mathcal{L}_{p\gamma \to pee}^p(\gamma, t) = 2 \frac{m_e}{m_p} \frac{\partial}{\partial \gamma}(\gamma \Gamma_p n_p(\gamma, t)) \quad (23)$$

where

$$\Gamma_p = \int_{2/\gamma}^{\infty} dx\, n_\gamma(x, t) \sigma_{pe}(x\gamma) \quad (24)$$

is the collision rate and $\sigma_{pe}(y)$ is the cross-section of the process in units of the Thomson cross-section as a function of the photon energy $y$ as seen in the rest frame of the proton.

On the other hand, the rate at which this process injects electrons and positrons, (we make no distinction between them) is

$$\mathcal{Q}_{p\gamma \to pee}^e(\gamma, t) = 2\Gamma_p n_p(\gamma, t). \quad (25)$$

### 3.3. Proton-photon pion production

To take this complicated process into account we assumed that the cross section is given by a $\delta$-function, i.e., $d\sigma_{p\pi}(y)/dy = \sigma_{p\pi}^0 \sigma_T \delta(y - y_0)$, where again $y$ is the photon energy seen from the proton rest frame, $\sigma_{p\pi}^0 = .25$ and $y_0 = 10^3$. We also take the inelasticity to be $k_p = .2$ and the multiplicity to be 1. These values are found to be in good agreement (better than 10%) with the results of Begelman et al. (1990) who calculate the proton losses due to photopion production using the full cross section. We consider the two basic channels

(a) $$p + \gamma \to p + \pi^0 \quad (26)$$

(b) $$p + \gamma \to n + \pi^+ \quad (27)$$

to be equally probable. Whereas the outgoing protons of channel (a) are effectively trapped, the neutrons of channel (b) are assumed to escape the source without further attenuation. Thus we treat channel (a) as an energy loss process which preserves proton number in contrast to channel (b) which is treated as a catastrophic proton loss. The photopion collisions of channel (a) move protons of Lorentz factors between $\gamma$ and $\gamma + d\gamma$ to lower ones and add photons to this range which, before interaction, had Lorentz factors around $\gamma' = \gamma/(1 - k_p)$. Thus, using the $\delta$-function approximation for the cross section, we find for channel (a):

$$\mathcal{L}_{p\gamma \to p\pi, a}^p(\gamma, t) =$$
$$\frac{\sigma_{p\pi}^0}{2} \left[ \frac{1}{\gamma'} n_p(\gamma', t) n_\gamma(y_0/\gamma', t) - \frac{1}{\gamma} n_p(\gamma, t) n_\gamma(y_0/\gamma, t) \right] \quad (28)$$

and for the catastrophic losses of channel (b):

$$\mathcal{L}_{p\gamma \to p\pi, b}^p(\gamma, t) = -\frac{\sigma_{p\pi}^0}{2} \frac{1}{\gamma} n_p(\gamma, t) n_\gamma(y_0/\gamma, t) \quad (29)$$

As neutral pions decay essentially instantaneously into $\gamma$-rays, channel (a) will provide a source term in the photon equation Eq. (13). Assuming the two $\gamma$-rays have equal energy (see, for example, Stecker 1968), this quantity ($x$) is related to the incoming proton energy $\gamma_1$ by $x = (m_p/2m_e)k_p\gamma_1$. Thus, we can write the photon source term as

$$\mathcal{Q}_{p\gamma \to p\pi}^\gamma(x, t) = n_p(\gamma_1) n_\gamma(x_1) \frac{\sigma_{p\pi}^0}{x} \quad (30)$$

where $x_1 = k_p m_p y_0/(2 x m_e)$.

Three neutrinos and a positron are created in the decay chain of a $\pi^+$ from channel (b). Assuming again that these are produced with equal energies, we can write the electron source term analogous to the photon term in Eq. (30):

$$\mathcal{Q}_{p\gamma \to p\pi}^e(\gamma, t) = \frac{1}{2} n_p(\gamma_2) n_\gamma(x_2) \frac{\sigma_{p\pi}^0}{\gamma} \quad (31)$$

where $\gamma_2 = 4\gamma m_e/(k_p m_p)$ and $x_2 = k_p m_p y_0/(4\gamma m_e)$.

Finally the neutrino emissivity can be written as

$$\mathcal{Q}_{p\gamma \to p\pi}^\nu(E_\nu, t) = \frac{3}{2} n_p(\gamma_3) n_\gamma(x_3) \sigma_{p\pi}^0 \frac{m_e c^2}{E_\nu} \quad (32)$$

where $\gamma_3 = 4E_\nu/(k_p m_p c^2)$ and $x_3 = k_p m_p c^2 y_0/(4E_\nu)$.

*3.4. Synchrotron radiation*

Classical synchrotron radiation is treated in several texts (e.g., Ginzburg & Syrovatskii 1965). Using our normalisation, the rate $\eta(x,\gamma)$ at which a single electron electron emits photons into the frequency range d$x$ is given by:

$$\eta(x,\gamma) = \frac{\sqrt{3}\alpha_f b \sin\theta}{x}\left(\frac{Rm_e c}{h}\right)F[2x/(3b\sin\theta\gamma^2)] \quad (33)$$

where $b = B/B_c$ denotes the magnetic field in units of the critical field $B_c = m_e^2 c^3/(e\hbar) = 4\cdot 414\times 10^{13}\,\text{G}$, $\alpha_f$ is the fine structure constant, $\theta$ the particle's pitch angle and we have used the standard notation of Ginzburg and Syrovatski for the function $F(x)$. Provided the energy of the emitted photon is much smaller than that of the emitting particle, the loss process can be considered continuous and represented by a first-order differential operator in the equation for electrons Eq. (11):

$$\mathcal{L}_{\text{syn}}^e = \frac{\partial}{\partial\gamma}\left(n_e(\gamma,t)\int_0^\infty dx\,\eta(x,\gamma)\right)$$
$$= \frac{4}{3}\ell_B \frac{\partial}{\partial\gamma}\left[\gamma^2 n_e(\gamma,t)\right]. \quad (34)$$

Here we have introduced the 'magnetic compactness':

$$\ell_B = \left(\frac{U_B}{m_e c^2}\right)\sigma_T R \quad (35)$$

in which $U_B = B^2/8\pi$ is the magnetic energy density and have replaced $\sin^2\theta$ by its average ($=2/3$) for isotropic electrons. Equation (34) contributes an amount $4\ell_B \gamma_{\min}^2 n_e(\gamma_{\min},t)/3$ to the source term $\mathcal{Q}^{e,\text{cool}}$ for cool electrons.

The source term in the photon equation can be found using the '$\delta$-function' approximation, originally introduced by Hoyle (1960) in which the emission of a single electron is approximated as monochromatic:

$$\eta(x,\gamma) \approx q_0 x_0 \delta(x-x_0), \quad (36)$$

and the quantities $q_0$ and $x_0$ must be determined by imposing requirements on the accuracy of the approximation. To reproduce the correct value for the total energy emitted by a single electron we find by integrating Eqs. (33) and (36) over $x$

$$q_0 x_0 = \frac{4}{3}\ell_B \gamma^2. \quad (37)$$

One further constraint could be obtained by requiring the production rate of photons to be given precisely by the delta-function approximation. This leads to $x_0 \approx 0\cdot 46\, b\gamma^2$. However, we prefer the simple expression:

$$x_0 = b\gamma^2 \quad (38)$$

which corresponds formally to requiring that the approximation yields an accurate value for the moment $\int dx\,\eta(x,\gamma)x^{-0.37}$. The appropriate photon source term follows simply from Eqs. (36) and (37):

$$\mathcal{Q}_{\text{syn}}^\gamma \approx \frac{4\ell_B}{3b}\int_1^\infty d\gamma\,n_e(\gamma,t)\delta(x-b\gamma^2)$$
$$= \frac{2}{3}\ell_B b^{-3/2} x^{-1/2} n_e(\sqrt{x/b},t) \quad (39)$$

*3.5. Synchrotron self-absorption*

Low energy electrons interact strongly with the radiation field if their Lorentz factor is such that the synchrotron photons they emit are reabsorbed within the source (McCray 1969, Ghisellini et al. 1988). The photon spectrum too is strongly affected below the self-absorption frequency (at which the optical depth equals unity). Such behaviour can be modelled using a second order derivative in momentum in the electron equation (Ghisellini et al. 1988). However, we do not wish to discuss in detail the heating and cooling processes of low energy electrons in this paper, nor are the details of the low energy spectrum important for our investigations. Consequently, we treat synchrotron self-absorption simply as a sink of soft photons.

Starting from standard formulas for the absorption coefficient $\alpha_\nu$ (e.g., Eq. (6.50) of Rybicki & Lightman 1979), and using the $\delta$-function approximation Eq. (36) one can readily derive

$$\mathcal{L}_{\text{ssa}}^\gamma \equiv R\alpha_\nu n_\gamma(x,t)$$
$$= \frac{\pi}{6\alpha_f}x^{-1/2} b^{-1/2} n_\gamma(x,t)\left[\frac{\partial}{\partial\gamma}\left(\frac{n_e}{\gamma^2}\right)\right]_{\gamma=(x/b)^{1/2}} \quad (40)$$

When only synchrotron emission and self-absorption are included, the photon kinetic equation (13) is:

$$\frac{\partial n_\gamma(x,t)}{\partial t} = \frac{2}{3}\ell_B b^{-3/2} x^{-1/2} n_e(\sqrt{x/b},t)$$
$$+ \frac{\pi}{6\alpha_f} b^{-1/2} x^{-1/2} n_\gamma(x,t)\left[\frac{\partial}{\partial\gamma}\left(\frac{n_e}{\gamma^2}\right)\right]_{\gamma=(x/b)^{1/2}} \quad (41)$$

For an equilibrium electron distribution $n_e \propto \exp(-\gamma/T)$ (where $T$ is the temperature in units of $m_e c^2/k_B$) the stationary photon distribution is the Rayleigh-Jeans distribution $n_\gamma^{\text{RJ}} = 4\pi R\sigma_T(mc/h)^3 Tx$, whereas for a power-law electron distribution one obtains a stationary self-absorbed spectrum $n_\gamma \propto x^{3/2}$. These results agree with those found from an exact treatment of synchrotron emission.

*3.6. Inverse Compton scattering*

There is a close analogy between Compton scattering and synchrotron radiation, as has been pointed out by Felten & Morrison (1966). In the case of synchrotron radiation, the classical treatment, in which the energy of the emitted photon is small compared to the energy of the electron, is adequate for our purposes. This is not so for Compton scattering, because events in the "Klein-Nishina regime" turn out to be important. We therefore divide our treatment of scattering into two parts.

For collisions occurring in the classical or Thomson regime we use a method closely similar to that used for synchrotron radiation. The electron loss term is:

$$\mathcal{L}_{\text{ics,T}}^e(\gamma,t) = \frac{4}{3}U_T \frac{\partial}{\partial\gamma}\left(\gamma^2 n_e(\gamma,t)\right) \quad (42)$$

where

$$U_T = \int_0^{x_T} dx'\,x' n_\gamma(x',t) \quad (43)$$

is the energy density of photons with which the electron interacts via an inverse Compton scattering in the Thomson regime i.e., photons of energy $x'$ less than $x_T \approx 3/(4\gamma)$. Equation (42) gives rise to an injection of cooled electrons at a rate $4U_T\gamma_{\min}^2 n_e(\gamma_{\min}, t)/3$. The photon source term in the Thomson regime is given by:

$$\mathcal{Q}_{\mathrm{ics,T}}^\gamma(x,t) = \frac{\sqrt{3}}{4}\int_0^{\min[3/(4x),3x/4]} dx' \\ x'^{-1/2}x^{-1/2}n_e(\sqrt{3x/4x'},t)n_\gamma(x',t) \quad (44)$$

Photons with $x > x_T$ will interact with electrons of energy $\gamma$ in Klein-Nishina scatterings. In order to treat this much more complicated case, we assume that an electron loses all its energy in one such collision and joins the population of cooled particles. Approximating the cross-section by $\sigma(x') \simeq \sigma_T/x'$ where $x' = \gamma x$, the electron loss term can be written

$$\mathcal{L}_{\mathrm{ics,KN}}^e(\gamma,t) = \frac{n_e(\gamma,t)}{\gamma}\int_{x_T}^\infty dx' \, \frac{n_\gamma(x',t)}{x'} \quad (45)$$

Because the scattered photons emerge with the energy of the incoming electron, the photon source term is simply

$$\mathcal{Q}_{\mathrm{ics,KN}}^e(x,t) = n_e(x,t)\int_{3/4x}^\infty dx' \, (xx')^{-1}n_\gamma(x',t) \quad (46)$$

Note that these are only approximate expressions since the logarithmic energy dependence of the cross section has been neglected. Nonetheless, this does not affect our results because rates in the Klein-Nishina regime are in any case suppressed by the proportionality of the cross section to the inverse of the energy (the "Klein-Nishina" cut-off).

### 3.7. Photon downscattering on cold electrons

Inverse Compton scattering describes the *upscattering* of photons by relativistic electrons. In addition, cool electrons ($\gamma \approx 1$), *downscatter* hard photons. The Thomson optical depth of cool electrons is in our normalisation simply $\tau_T = N_e^{\mathrm{cool}}$. The rate of photon downscattering in the Thomson regime can then be found from the Kompaneets equation, assuming the temperature of cool electrons is zero and neglecting the stimulated scattering term:

$$\mathcal{L}_{\mathrm{cs}}^\gamma(x,t) = \tau_T\frac{\partial}{\partial x}\left[\langle\Delta x\rangle n_\gamma(x,t)\right]. \quad (47)$$

$\langle\Delta x\rangle$ is the average energy shift of a photon with energy $x$ colliding with an electron at rest. This quantity was calculated from the relation

$$\langle\Delta x\rangle = \frac{\int_{x'_{\min}}^{x'_{\max}} dx'(x-x')d\sigma(x,x')/d\Omega}{\int_{x'_{\min}}^{x'_{\max}} dx'd\sigma(x,x')/d\Omega}, \quad (48)$$

where $d\sigma(x,x')/d\Omega$ is the differential cross section for Compton scattering (see, for example, Akhiezer & Berestetskii 1969) and the limits of the integration $x'_{\max} = x$ and $x'_{\min} = x/(2x+1)$ are determined from kinematical constraints. For $x \ll 1$, $\langle\Delta x\rangle = x^2/(2x+1) \simeq x^2$, which when substituted in Eq. (47) yields the usual Kompaneets equation (under the simplifications assumed here). For $x \lesssim 1$, $\langle\Delta x\rangle$ departs from the $x^2$ dependence, as relativistic effects start becoming important. In order to take into account the spatial diffusion of photons inside the source, we follow Lightman & Zdziarski (1987) and write for the photon escape time

$$t_{\gamma\mathrm{esc}} = 1 + \frac{\tau_T}{3}f(x), \quad (49)$$

where

$$f(x) = \begin{cases} 1 & x \leq 0.1 \\ (1-x)/0.9 & 0.1 < x < 1 \\ 0 & x \geq 1 \end{cases} \quad (50)$$

Therefore, when only the effects of photon downcomptonization are included, the photon kinetic equation (13) becomes

$$\frac{\partial n_\gamma(x,t)}{\partial t} + \frac{n_\gamma(x,t)}{1+\frac{1}{3}\tau_T f(x)} = \tau_T\frac{\partial}{\partial x}\left(\langle\Delta x\rangle n_\gamma(x,t)\right). \quad (51)$$

For $x \geq 1$, Eq. (47) is no longer valid as the photons lose a large fraction of their energy in each collision. Compton scattering can then be considered as a catastrophic loss process and to deal with this case, we follow Svensson (1987) and write

$$\mathcal{L}_{\mathrm{cs}}^\gamma(x,t) = \tau_T n_\gamma(x,t)/x. \quad (52)$$

The corresponding injection of (relativistic) electrons is

$$\mathcal{Q}_{\mathrm{cs}}^e(\gamma,t) = \tau_T n_\gamma(\gamma,t)/\gamma. \quad (53)$$

This injection comes from the upscattering of cooled electrons, and when integrated over all energies contributes to $\mathcal{L}^{e,\mathrm{cool}}$ in Eq. (12).

### 3.8. Photon-photon pair production

This process acts as a sink of high energy photons, as well as an injection term of electrons. To calculate the loss rate of photons we follow Coppi and Blandford (1990):

Photons of energy $x$ are lost at a rate

$$\mathcal{L}_{\gamma\gamma\rightarrow ee}^\gamma(x,t) = n_\gamma(x,t)\int_0^\infty dx' \, n_\gamma(x',t)R_{\gamma\gamma}(xx') \quad (54)$$

where $R_{\gamma\gamma}$ is a fit to the reaction rate given by

$$R_{\gamma\gamma}(\omega) = 0\cdot 652\frac{\omega^2-1}{\omega^3}\ln(\omega)H(\omega-1) \quad (55)$$

where $H(y)$ is the Heaviside function.

Photon-photon pair production is also responsible for the injection of relativistic electrons and positrons. Assuming, as in the case of Bethe-Heitler pair production, that the electron and positron emerge with equal energy $\gamma$, and noting that the photon-photon pair production process requires at least one hard photon of energy $x > 1$, which interacts predominantly with a soft photon of energy around $1/x$, we find from conservation of energy

$$2\gamma = x + \frac{1}{x}$$
$$\approx x. \quad (56)$$

The injection term for electrons is then

$$\mathcal{Q}_{\gamma\gamma\rightarrow ee}^e(\gamma,t) = 4n_\gamma(2\gamma,t)\int_0^\infty dx' \, n_\gamma(x',t)R_{\gamma\gamma}(2\gamma x'). \quad (57)$$

*3.9. Electron-positron pair annihilation*

This process is the inverse of photon-photon pair production described above. It acts as a sink of electrons and positrons and a source for photons. In this treatment we will consider only the "cosmic-ray" case discussed by Svensson (1982) and Aharonian et al. (1983) in which relativistic electrons/positrons annihilate only on the cool particles and not amongst themselves. With this assumption, the losses of electron-positron pairs can be written

$$\mathcal{L}^e_{ee \to \gamma\gamma}(\gamma, t) = \tau_T R_{ann}(\gamma) n_e(\gamma, t) \tag{58}$$

where $R_{ann}$ is given by Coppi & Blandford (1990)

$$R_{ann} = \frac{3}{8\gamma} \left[ \gamma^{-1/2} + \ln \gamma \right]. \tag{59}$$

Photons produced by pair annihilation have $\langle x \rangle = (\gamma + 1)/2 \simeq \gamma/2$ so the photon term can be written

$$\mathcal{Q}^\gamma_{ee \to \gamma\gamma}(x, t) = 2\tau_T R_{ann}(2\gamma) n_e(2\gamma, t). \tag{60}$$

Finally the rate of Eq. (58) integrated over energy gives us the rate of cooled electron removal due to electron-positron annihilation, it contributes thus to $\mathcal{L}^{e,cool}$ in Eq. (12). Since this rate does not include the annihilation of cooled pairs among themselves, we use an extra term

$$\mathcal{L}^{e,cool}(t) = -\frac{3}{4} \left[ N_e^{cool}(t) \right]^2. \tag{61}$$

Furthermore, we assume that the rate of emission of the 511 keV annihilation rate is given by the above equation, so we write

$$\mathcal{Q}^\gamma_{ee \to \gamma\gamma}(1, t) = \frac{3}{4} \left[ N_e^{cool}(t) \right]^2. \tag{62}$$

## 4. The numerical method

*4.1. Discretization*

When the physical process described in Sect. 3 are included, the system of kinetic equations (6), (11) and (13) is an integro-differential set for the distributions $n_p(\gamma, t)$, $n_e(\gamma, t)$ and $n_\gamma(x, t)$. Our approach to the numerical solution is to discretize the variables $\gamma$ and $x$, and to integrate the resulting (stiff) set of coupled ordinary differential equations forwards in time. Because we use crude but computationally rapid approximations to the physical processes we are able to integrate the equations using a standard NAG-library routine for stiff systems.

The grid chosen for discretization is equally spaced in the logarithm of $\gamma$, the same grid being used for both protons and electrons. In order to accommodate a large dynamic range of the distributions $n_p(\gamma_{max}, t)$, $n_e(\gamma_{max}, t)$ and $n_\gamma(x, t)$ we use their logarithms in the numerical integration. In a typical run the resolution is 10 bins per decade for protons and electrons. This is a rather coarse grid; however, we found runs with higher resolutions to be excessively time consuming. The highest grid point $\gamma_{max}$ is adjusted so that the adjacent bins to $n_e(\gamma_{max}, t)$ and $n_p(\gamma_{max}, t)$ remain negligibly small throughout the run. We find a value of $\gamma_{max} = 10^9$ is adequate to ensure this for the models presented here. We choose the lowest grid point to be at $\gamma_{min} = 10^{0.1}$. Electrons which cool through this boundary join the population of cool electrons.

For the photons, the grid spacing in the logarithm of $x$ is chosen to be twice as large as that in $\log \gamma$ for protons (or electrons). Because the $\delta$-function approximation is used in the synchrotron source term, Eq. (39), this grid spacing eliminates the need for interpolation; each electron grid point is associated with a single photon grid point. Thus, since the softest photons are produced by electron synchrotron radiation, we take $x_{min} = b\gamma_{min}^2$. However, inclusion of the pion producing processes requires the photon grid to be extended above the $x$ corresponding to the synchrotron radiation of electrons of $\gamma_{max}$, because of the very hard photons produced in $\pi^0$-decay. We therefore choose $x_{max} = \gamma_{max}$ and check that the photon bins adjacent to $x_{max}$ remain practically empty throughout the run.

Integrals over the photon and electron distributions are straightforwardly converted into summations using the trapezoidal rule, but care must be taken in discretizing the first order derivatives with respect to $\gamma$. In the proton equation (6) and in the electron equation (11) these terms describe 'continuous' acceleration and losses and in order to avoid a numerical instability, an upstream difference must be taken. For particles undergoing losses 'upstream' means larger $\gamma$, whereas for the accelerating particles it means smaller $\gamma$. Thus, for electrons, which experience no acceleration, the value of the distribution at $\gamma_{max}$ is held constant at a negligibly small value. The value at $\gamma_{min}$ is computed. On the other hand, protons experience only acceleration at $\gamma_{min}$, so that $n_p(\gamma_{min})$ is held constant and is related to the effective injection rate. At the upper end of the scale, losses always dominate for protons of $\gamma_{max}$, so that, just as for electrons, $n_p(\gamma_{max})$ is held constant as a boundary condition. Similarly, in order to take into account the first order derivative with respect to the photon energy $x$ (Eq. 47), we take an upstream difference while $n_\gamma(x_{max})$ is held constant at a negligibly small value.

*4.2. Performance Checks*

In order to test both the code and the treatment of the physical processes, we have performed various runs for cases in which either an analytic solution is available, or in which we can compare our results with calculations in the literature. In the following sections, when appropriate, we will refer to the electron injection compactness, defined in our normalisation by

$$\ell_e = \frac{1}{3} \int_{\gamma_{min}}^{\gamma_{max}} d\gamma \, (\gamma - 1) Q_e(\gamma) \tag{63}$$

where

$$\begin{aligned} Q_e(\gamma) &= Q_{e,0} \gamma^{-s} \quad \text{for } \gamma_{min} < \gamma < \gamma_{max} \\ &= 0 \quad \text{otherwise} \end{aligned} \tag{64}$$

is an external electron injection rate introduced for test purposes.

4.2.1. Proton acceleration

For the first test of the code we check the way the numerical method deals with proton acceleration. As stated above, protons are effectively injected at a Lorentz factor $\gamma_{inj} = 10^{0.1}$. The solution of Eq. (1) in the case where $\mathcal{L}^p = 0$ is given by Eq. (2). Figure (1) shows the evolution of such a proton spectrum in the case where $t_{\rm acc} = t_{\rm esc} = 3 t_{\rm cross}$ at times $t = 10$, 20, 30 and 40 $t_{\rm cross}$ (full lines). The dotted lines represent the analytical solution for this case. As can seen from this figure, the adopted numerical scheme is fairly accurate in tracing the time evolution of the cut-off momentum. It also reproduces very precisely the expected power law, which has a slope $-2$. Some numerical diffusion is unavoidably introduced, which po-tentially influences the shape of any sharp features in the photon spectrum. However, our acceleration model is in any case not accurate in the region of the spectral cut-off – a finite time-lag between scatterings as well as finite (but small) jumps in the particle energy (e.g., on crossing a shock) both lead to a smearing out of the sharp cut-off.

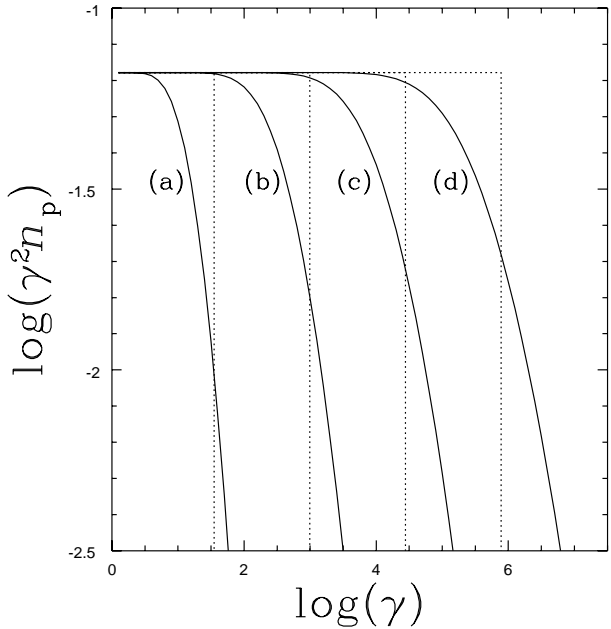

**Fig. 1.** The evolution of the proton spectrum in the no-loss case with $t_{\rm acc} = t_{\rm esc} = 3$ (full lines) at instants $t = $ (a) 10, (b) 20, (c) 30 and (d) 40 (all times are given in units of $t_{\rm cross}$). The dotted lines represent the corresponding analytical solutions.

#### 4.2.2. Synchrotron radiation & synchrotron self-absorption

To check our treatment of synchrotron radiation we inject electrons with a power law spectrum at a constant rate given by Eq. (64) and include only the processes of synchrotron radiation, synchrotron self-absorption and photon escape. The resulting electron distribution can be found analytically (Kardashev 1962). For $s > 1$ and $\gamma > \gamma_{\rm min}$, there are three regimes. Defining the break point by

$$\gamma_{\rm br}(t) = \gamma_{\rm max}/(1 + 4\gamma_{\rm max}\ell_{\rm B} t/3) \qquad (65)$$

we have:

1. For $\gamma > \gamma_{\rm max}$,
$$n_{\rm e}(\gamma,t) = 0 \quad \text{for all } t \qquad (66)$$

2. At early times i.e., for $\gamma < \gamma_{\rm br}$, the spectrum is
$$n_{\rm e}(\gamma,t) = \frac{3Q_{e,0}\gamma^{-s}}{4\ell_{\rm B}(s-1)}\left[\frac{1}{\gamma} - \frac{1}{\gamma}(1 - 4\ell_{\rm B}\gamma t/3)^{s-1}\right] \qquad (67)$$
which, for $4\ell_{\rm B}\gamma t/3 \ll 1$, is a power law of index $s$ in $\gamma$ with an amplitude which increases linearly with time: $n_{\rm e} \approx Q_{e,0} t \gamma^{-s}$.

3. For $\gamma > \gamma_{\rm br}$, electrons have had time to cool, and the spectrum steepens:
$$n_{\rm e}(\gamma,t) = \frac{3Q_{e,0}}{4\ell_{\rm B}(s-1)\gamma^2}\left(\gamma^{-(s-1)} - \gamma_{\rm max}^{-(s-1)}\right) \qquad (68)$$

Figures (2A) and (2B) show the numerically calculated electron and photon spectra at times $t = 10^{-4}, 10^{-3}, 10^{-2}, 10^{-1}$ and 10 (in units of $t_{\rm cross}$), choosing $b = 10^{-6}$, $\ell_{\rm B} = 0.075$, $s = 2$, $\gamma_{\rm min} = 3$ and $\gamma_{\rm max} = 3.10^5$. The electron spectra show breaks at energies which agree well with Eq. (65). For $\gamma \ll \gamma_{\rm br}$ the electron distribution is approximately a power-law $n_{\rm e} \propto \gamma^{-\alpha}$, has the same index as the injection term ($\alpha = s$) and increases linearly with time, whereas for $\gamma \gg \gamma_{\rm br}$ electrons have cooled giving a time-independent spectrum of index $\alpha = s + 1 = 3$.

The photon spectrum shows similar behaviour. Here, however, there are two breaks. The one at high energies which we shall call $x_{\rm br}$ is related to $\gamma_{\rm br}$ by $x_{\rm br} = b\gamma_{\rm br}^2$; that at low energies, called $x_{\rm ssa}$, is caused by synchrotron self-absorption. Photons with $x_{\rm ssa} \ll x \ll x_{\rm br}$ have a power-law distribution $n_\gamma \propto x^{-p}$ with $p = 1.5$, in agreement with the standard formula $p = (\alpha + 1)/2$. Photons with $x > x_{\rm br}$ correspond to the 'cooled' part of the electron distribution ($\gamma > \gamma_{\rm br}$) and have a power-law of index of $p = 2$ again in agreement with the standard formula (in this case with $\alpha = s + 1 = 3$). Photons with $x < x_{\rm ssa}$ show the characteristic self-absorbed distribution $n_\gamma \propto x^{3/2}$ which, as can be seen from Eq. (41), is produced by any power-law electron distribution, independent of the value of $s$. It is interesting to note that while $x_{\rm br}$ decreases with time (since $\gamma_{\rm br}$ also decreases), $x_{\rm ssa}$ increases. This happens because as the system evolves in time, the number density of low energy electrons and, hence, the optical depth to self-absorption increases.

#### 4.2.3. Inverse Compton Scattering

A similar approach to the above can be taken for the process of inverse Compton scattering. We take a constant electron injection given by Eq. (64) and assume the scattering is with photons of an external black-body field of temperature $\Theta = kT/m_e c^2$ and compactness $\ell_{\rm bb}$. The external photon source can be written

$$n_{\rm bb} = \frac{45\ell_{\rm bb}}{\pi^4 \Theta^4} \frac{x^2}{e^{x/\Theta} - 1}. \qquad (69)$$

We then solve numerically the kinetic equation for electrons retaining only the Compton loss terms, and the kinetic equation for photons including inverse Compton scattering and escape from the boundary. In order to illustrate the effects of the transition from the Thomson to the Klein-Nishina regimes (which, in our case, are described by two different electron loss and photon gain rates) we solve the equations keeping all parameters the same and increasing only the upper limit of the electron injection $\gamma_{\rm max}$. In the particular case considered here we choose $\Theta = 10^{-5}$ and $\ell_{\rm bb} = 1/3$ for the parameters of the external photon field, and $s = 2$, $\gamma_{\rm min} = 3$ and $Q_{e,0} = 10^{-5}$ for the electron injection parameters. The results are presented in Fig. (3) for $\gamma_{max}$ of (a) $3.10^3$, (b) $3.10^4$, (c) $3.10^5$, (d) $3.10^6$ and (e) $3.10^7$. Since the blackbody photons can be approximated as having an energy $\langle x_{bb}\rangle \simeq 3\Theta = 3.10^{-5}$, we have that for cases (a) and (b) all scatterings occur in the Thomson regime, whereas for the other cases, the more energetic electrons scatter in the Klein Nishina regime. Figure (3A) shows the *steady state* electron distribution and Fig. (3B) the *steady state* photon distribution. For cases (a) and (b) one obtains, just as in the case of synchrotron cooling, a power-law electron distribution of index $\alpha = s + 1 = 3$ and, correspondingly, a photon spectrum of index $p = 2$. For the other cases scattering in the Klein-Nishina regime is important and the electron spectrum becomes harder for $\gamma > 10^5$. In accordance with previous work (Jones 1994, Blumenthal and Gould 1970, Blumenthal 1971, Zdziarski 1989) we find an index $\alpha \simeq s - 1 = 1$ as $\gamma_{max}$ becomes large. However, as can be seen from Fig. (3B), the photon distribution does not break but retains its $p = 2$ slope throughout. This is to be expected for the special case of $s = 2$, because in the Klein Nishina case the photon distribution has a slope $p = \alpha + 1 = 2$, i.e., the radiated photons have essentially the injected electron index (Blumenthal & Gould 1970) and in the Thomson regime one has $p = (\alpha + 1)/2 = 2$. Although our approximate treatment reproduces the above properties quite well, the abrupt switch we use between the two scattering regimes produces artificially sharp breaks, such as can be seen in Fig. (3A).

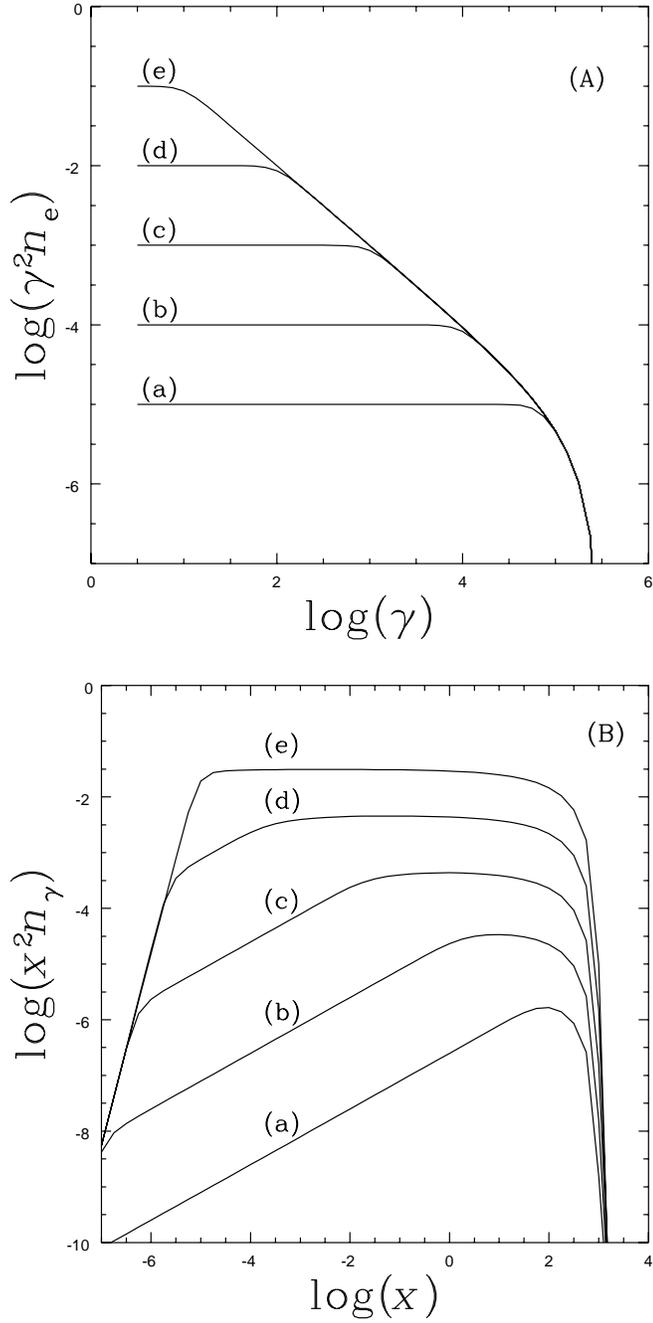

Fig. 2. (A) the electron distribution for synchrotron cooling and constant power-law injection. (B) the corresponding photon distribution including escape and synchrotron self-absorption. Here $b = 10^{-6}$, $\ell_B = 0.075$, $s = 2$, $\gamma_{min} = 3$ and $\gamma_{max} = 3.10^5$. The curves show the evolution of the electron and photon distributions towards steady state. Curve (a) is $t = 10^{-4}$ after electron injection, (b) is for $t = 10^{-3}$, (c) is for $t = 10^{-2}$, (d) is for $t = 10^{-1}$ and (e) is for $t = 10$. All times are expressed in units of $t_{cross}$.

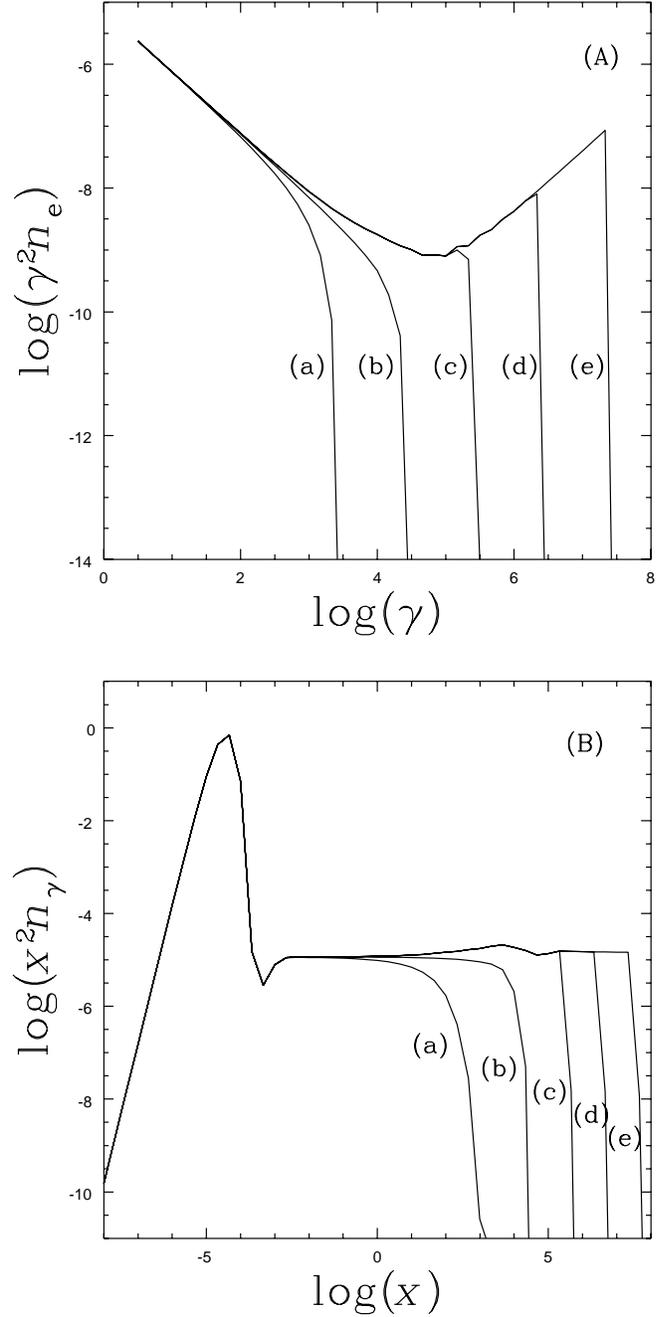

Fig. 3. Electron steady state and photon spectra in the case where the electron cooling is provided by inverse Compton scattering on thermal photons of temperature $\Theta = 10^{-5}$ and compactness $\ell_{bb} = 1/3$. The electrons were assumed to be injected with a power law of slope s=2 between $\gamma_{min} = 3$ and $\gamma_{max}=$ (a) $3.10^3$, (b) $3.10^4$, (c) $3.10^5$, (d) $3.10^6$ and (e) $3.10^7$.

#### 4.2.4. Electromagnetic cascades

As a test for Compton scattering, photon-photon pair production, electron-positron annihilation and Compton downscattering, we present a series of runs showing pair cascades, which have been investigated by many authors [see Svensson (1989) for a review]. Electrons injected with $s = 1.5$, $\gamma_{min} = 3$ and $\gamma_{max} = 7.10^3$ are allowed to cool on an external photon field with $\Theta = 10^{-5}$ and $\ell_{bb} = \ell_e$. For electron compactnesses $\ell_e$ of (a) $1/(4\pi)$, (b) $30/(4\pi)$ and (c) $1000/(4\pi)$ we ran the code until a steady state was reached. The electrons cool on the external photons as in the case examined in Sect. (4.2.3). However, especially when the compactness is high, the resulting $\gamma$-rays do not escape freely. Instead, they produce electron-positron pairs upon interacting with soft photons. These pairs, in their turn, produce more $\gamma$-rays, initiating an electromagnetic cascade. The computed photon spectra are shown in Fig. (4). In case (a), where the compactness is quite low, the computed spectrum is the same as that expected in the pure inverse Compton scattering case, i.e., there is no pair feedback. For intermediate compactness (case b) there is some pair production and the spectrum steepens because of the redistribution of the luminosity to lower energies by pair production. Finally for high compactness (case c) there is substantial pair production and $\gamma$-rays of $x > 1$ are effectively absorbed. However, the steepening of the spectrum at energies $x < 1$ is due not to pair production, but to the downscattering of photons on cooled pairs – an effect which is important for $1/\tau_T^2 < x < 1$. We have compared our results with those found by Lightman and Zdziarski (1987), for the same initial conditions (their Fig. 3 – note that there is a $4\pi$ difference in the definition of compactness between Lightman and Zdziarski and the present work). Overall we found a very good agreement, the only major difference being in case (c) around $x \sim 10^{-4}$. This effect can be attributed to the upscattering of soft photons by thermal electrons – a process we neglect.

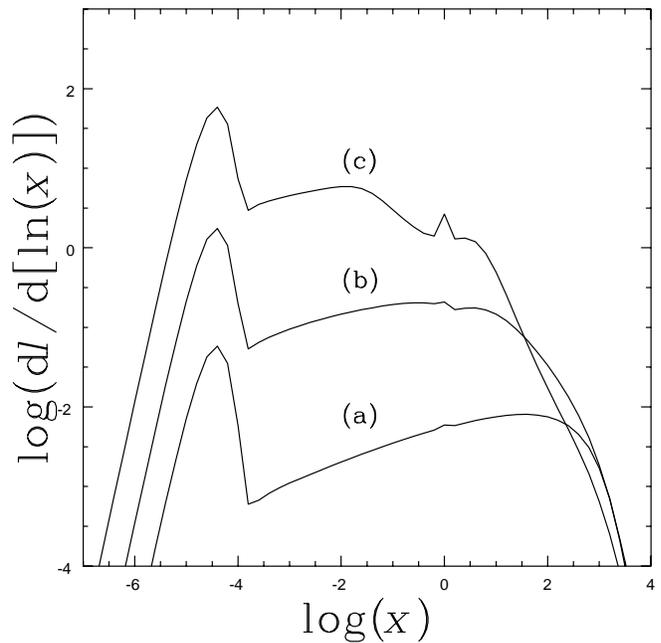

**Fig. 4.** Photon spectra obtained from electromagnetic cascades. The injected electrons have a slope $s = 1.5$ between $\gamma_{min} = 3$ and $\gamma_{max} = 7.10^3$ and they cool on an external photon field with $\Theta = 10^{-5}$ and $\ell_{bb} = \ell_e$. Curve (a) corresponds to an electron compactness $\ell_e$ of $1/(4\pi)$, curve (b) to $\ell_e = 30/(4\pi)$ and curve (c) to $\ell_e = 1000/(4\pi)$.

### 5. The Pair-Production–Synchrotron Instability

We turn now to an example of the use of the full code. The system of the kinetic equations (1), (11) & (13), is subject to various feedback effects and in this section we present one of them, namely the 'pair-production synchrotron instability' (PPS). This feedback effect manifests itself when the number density of relativistic protons with energy above a threshold $[\gamma \geq \gamma_{crit} = (2/b)^{1/3}]$ becomes larger than a certain critical value. In this case the relativistic protons produce pairs which subsequently radiate synchrotron photons energetic enough to provide the next generation of targets for the relativistic protons. This eventually leads to a runaway – (KM92).

The input parameters for the simulation are

1. the strength of the magnetic field, expressed as the ratio $b = B/B_c$. Although the threshold $\gamma_{crit}$ depends only weakly on $b$, the growth rate of the instability is more strongly affected. In the simulation, we choose for convenience a value of $4.4 \times 10^4$ G ($b = 10^{-9}$), rather higher than those normally estimated for AGNs.
2. the magnetic compactness $\ell_B$, which, given the magnetic field, determines the size of the emission region via Eq. (35). For this particular run we choose a size of $1.5 \times 10^{14}$ cm, corresponding to a light crossing time of about one hour. Such a value is appropriate for a rapidly variable Seyfert galaxy.

3. the acceleration time $t_{acc}$. The pair-production synchrotron instability threshold can be derived assuming acceleration is a slower process than all others (except escape, which has exactly the same timescale in our case, leading to a $n_p \propto \gamma^{-2}$ spectrum). Our treatment becomes inconsistent when the acceleration time is less than a light crossing time, since this implies small regions within the source where acceleration and losses compete, whereas we treat each of these processes as spatially homogeneous. Thus, we choose an acceleration time $t_{acc}$ equal to three light crossing times.

4. the compactness of the total nonthermal luminosity $\ell_{tot}$, which is related to the quantity $n_0$ defined in Eq. (8) by the relation (9). This parameter determines whether or not instability occurs. From KM92, we have for the threshold value $n_{crit}$:

$$n_{crit} = \frac{1}{3}b^{1/3}\left[\int_0^\infty dy\, \sigma_{pe}(y) y^{-5/3}\right]^{-1}. \quad (70)$$

where $\sigma_{pe}$ is the pair production cross section (in units of $\sigma_T$). For $n_0 > n_{crit}$ there exists a critical Lorentz factor $\gamma_{max}$ for the upper cut-off of the proton distribution, above which the system goes unstable. In the simulation we choose $n_0 = 1.1 n_{crit}$ and the instability takes off soon after the the maximum Lorentz factor exceeds the value $\gamma_{crit} = (2/b)^{1/3} = 1.3 \times 10^3$.

5. the background density of protons. We assume here that proton acceleration takes place in an environment where the thermal plasma density can be considered negligible, as may be the case in an accretion disk corona. Thus the only proton-proton interactions we consider are those of the relativistic protons amongst themselves.

Fig. (5) shows the evolution of the photon compactness (solid line) and also a quantity $d_p$ which we call the 'proton column-depth' and which gives a dimensionless measure of the energy contained in relativistic protons in the source:

$$d_p = \frac{m_p}{3 m_e} \int_{\gamma_{crit}}^\infty d\gamma\, \gamma n_p. \quad (71)$$

Initially, a few seed photons are produced by proton-proton collisions (i.e., in $\gamma$-rays from neutral pions and synchrotron photons from the electrons/positrons injected from the decay of charged pions – cf. Section 3.1). However, at about 20 crossing times the upper cut-off of the proton distribution becomes larger than $\gamma_{crit}$. The PPS instability then takes over and results in a rapid increase in the photon density at about 25 crossing times. The protons react to this growth after only a few crossing times, their energy density being reduced as a result of cooling on the produced photons. Because the pairs produced in this process cool rapidly, the photon density decreases on a time scale of about the light crossing time. This enables the acceleration process to act again, increasing the proton 'column-depth' until pair production restores the photon density. The resulting oscillations are damped for these parameters, and the the situation reaches a stationary state after several crossing times. The photon compactness in the stationary state is only roughly 50% of the total compactness $\ell_{tot}$ (shown by the dashed line). The remaining nonthermal power escapes the system in the form of either neutrinos or neutrons. The interrelation between $\ell_\gamma$ and $d_p$ is displayed in Fig. (6).

Figure (7A) shows the proton spectrum at five values of $t$ chosen in such a way as to show the onset and saturation of the instability. These snapshots are for $t_i = 23, 28, 33, 38$ and 43 ($i = 1 \ldots 5$) crossing times after the run was started. The corresponding photon distribution is shown in Fig. (7B). The photon compactness at these five instants is $\ell_\gamma \simeq 2, 4, 13, 28$ and 21 respectively. At $t_1$ (at which point the photons have still a low compactness – $\ell_\gamma \simeq 2$, and have not yet become important for proton losses) the protons have a power law distribution up to $\gamma_{\max}$ ($\simeq 2 \times 10^3$ in this particular case), i.e. at this stage their upper cut-off has just moved at values higher than $\gamma_{\text{crit}}(\simeq 10^3)$. The proton slope is $-2.02$, a little steeper than the canonical value of $-2$ and this can be attributed to the effects of proton-proton losses – cf. Section 3.1. At $t_2$ and $t_3$ the proton energy continues to increase with the photon energy density also increasing but still too low for effective proton losses. However, as the photon energy density increases (for $t \geq t_4$) losses become important and the protons cool down to values close to $3.10^4$, not too far from their final steady state distribution.

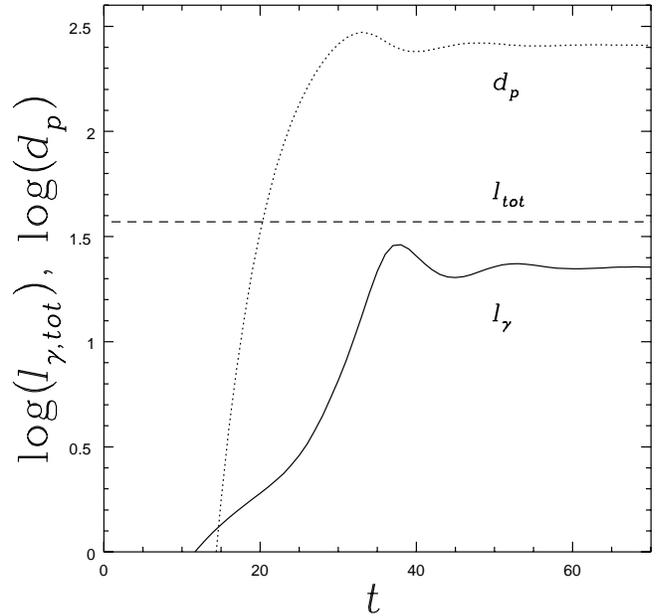

Fig. 5. Photon compactness (full line) and proton column-depth (dotted line) versus time expressed in units of crossing time in the case where the pair/synchrotron instability operates. Injection parameters are $n_0 = 1.1 n_{crit}$, $t_{acc} = 3 t_{cross}$ and $b = 10^{-9}$. The total compactness $\ell_{tot}$ [see Eq. (10)] is shown by the dashed line.

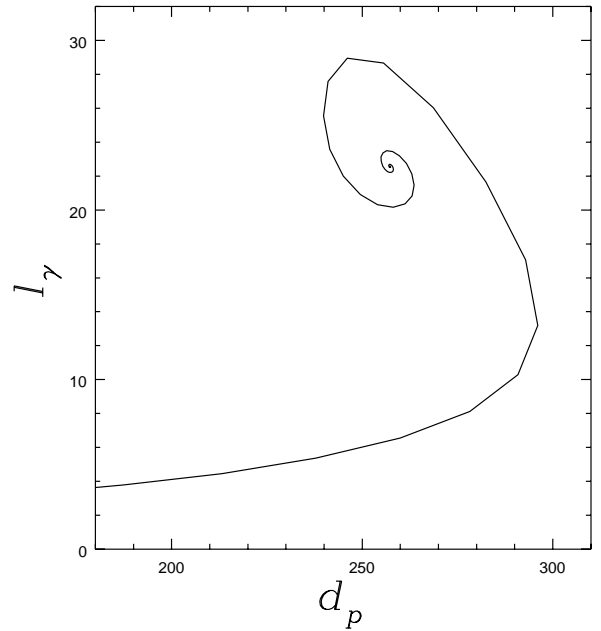

Fig. 6. Photon compactness versus proton column-depth for the injection parameters specified in Figure 5 and in the text.

On the other hand the photon spectrum has the following characteristics: As the instability manifests itself the photons have a spectral index of $-1.8$, close to the value of $-5/3$ predicted in the linear theory – see KM92. At high frequencies it shows a turnover which is given approximately by the relation $x_{\text{br}} \simeq b\gamma_{\max}^2$ – see sections (3.2) and (3.4). As $\gamma_{\max}$ moves to higher energies (i.e., for $t_1$ to $t_3$), so does $x_{\text{br}}$. However, as the protons cool (for $t = t_4$), $x_{\text{br}}$ moves to lower energies. We find that this break varies between 50 - 200 keV as it is to be expected from the formula for $x_{\text{br}}$ used above. At low frequencies, the photon spectrum shows strong synchrotron self-absorption as expected from such a compact source. Furthermore, as the source flares up, the self-absorption frequency increases. Again this is to be expected since during the linear stage of the instability the number density of electrons in the system increases.

Fig (8) depicts the photon *number* spectral index $a$ over the range 2-10 keV as a function of the photon compactness $\ell_\gamma$. The slope is a little steeper than the analytically calculated slope of $-5/3$ (KM92). It is worth noting that the spectrum becomes steeper as the photon luminosity decreases after reaching the peak.

The dimensionless neutrino emissivity $Q_\nu$ once steady state has been established is shown in Fig. (9). This is calculated using Eqs. (19) and (32). The neutrinos produced in pp collisions dominate at relatively lower energies ($E_\nu < 100 \, \text{GeV}$) and have a slope of $\simeq -2$, reflecting the parent proton distribution. The neutrinos produced in p$\gamma$ collisions dominate at higher energies, however they fall abruptly above a few TeV.

In this example the evolution of a system follows closely that predicted for the PPS instability. Out of all the processes described in Section 3 the leading ones are proton-photon pair production and synchrotron radiation. The effects of other processes such as photopion production, inverse Compton scattering and photon-photon absorption remain negligible throughout.

## 6. Summary

In this paper we develop a technique suitable for the investigation of AGN models in which the basic source of luminosity is the acceleration of nonthermal protons. Once these have reached sufficiently high energy, they are responsible for the injection of electron/positron pairs and photons. Such models have been suggested some time ago (Protheroe and

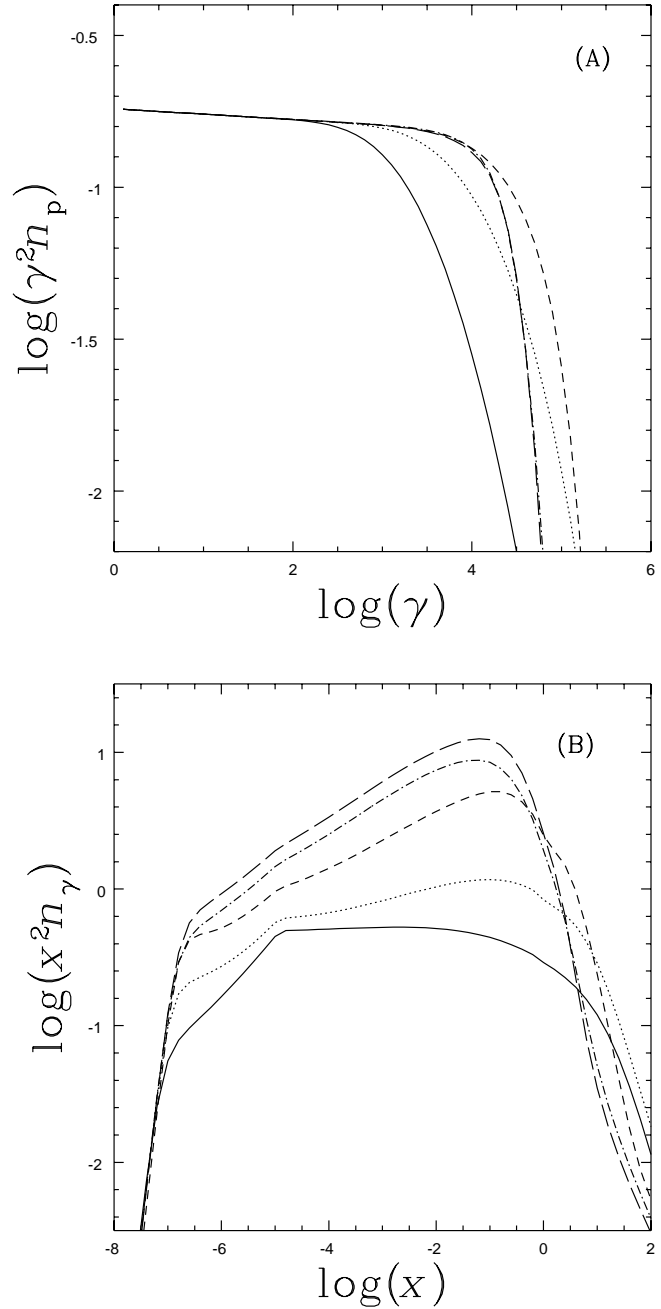

Fig. 7. (A) The proton distribution $n_p$ at five different times for the injection parameters specified in Figure 5. The solid line is for $t = 23$, the dotted line for $t = 28$, the short-dashed line for $t = 33$, the long-dashed line for $t = 38$ and the dot-dashed line for $t = 43$. All times are expressed in units of $t_{\text{cross}}$. (B) The corresponding photon spectrum.

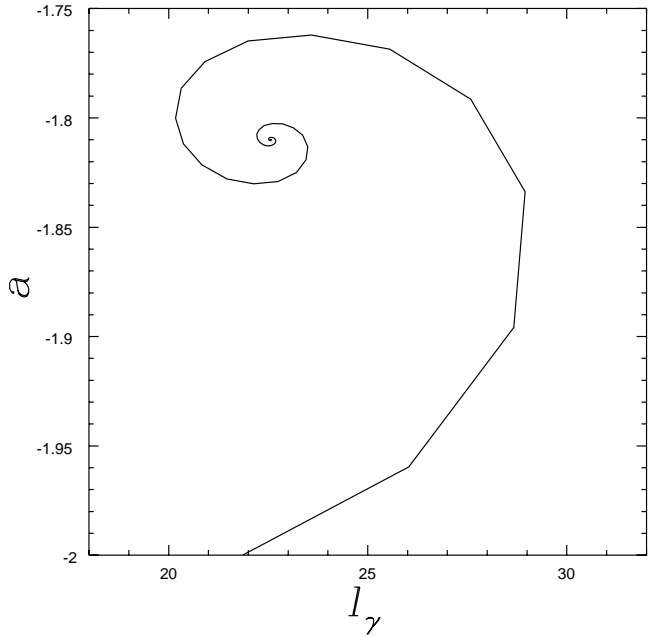

Fig. 8. The 2-10 keV photon number index $a$ versus photon compactness for the parameters given in Figure 5 and in the text.

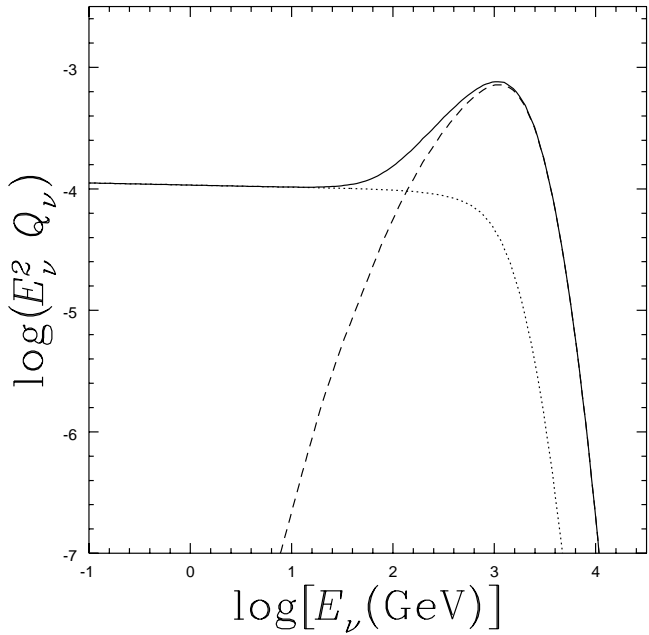

Fig. 9. The dimensionless neutrino emissivity $Q_\nu$ as a function of the neutrino energy in GeV once steady state has been established. The dotted line corresponds to the emissivity due to proton-proton collisions, the dashed line is the emissivity due to proton-photon collisions and the full line is the total emissivity. The injection parameters are as given in Figure 5 and in the text. The true emissivity (number of neutrinos emitted per second per unit energy per unit source volume) is $a_\nu = Q_\nu/(m_p c R^2 \sigma_T)$

Kazanas 1983, Kazanas and Ellison 1986), but the new aspect of our work is the self-consistent treatment of the interactions between the three important components of the model: protons, electrons/positrons and photons.

To achieve this, we follow the kinetic equation approach using spatially averaged distributions and an 'escape probability' formalism. Self-consistency is achieved by including source and loss terms in these equations which account for all of the important interactions between the components. Furthermore, we avoid prescribing the injection of nonthermal power by using an explicit model for the acceleration of the protons. This results in a rather constrained parameter space for our problem. The system is fully specified when the details of proton acceleration and the injection of low energy seed particles are determined. The electron and photon equations contain no additional free parameters. Thus, the external parameters usually employed in pair plasma problems, such as the energy with which nonthermal pairs or photons are injected and their compactnesses (see Lightman and Zdziarski 1987 and Svensson 1987) are in our case absent.

The resulting system is sufficiently complicated that a number of checks of both the physically motivated approximations used for the elementary processes, as well as the numerical method, is essential. These checks indicate that the method provides a useful tool for more detailed applications to AGN. As well as discussing tests, we present in this paper one example of results which are obtained with all processes included – one in which the feedback loop of the the pair/synchrotron instability is expected to operate. We find this instability is indeed present, and is not significantly suppressed by processes such as proton-proton interaction or photo-pion production. The prediction of an X-ray spectral index of $-5/3$ which stems from a linear analysis of this instability (KM92) is shown to be unaffected by the additional processes we include, and to extend well into the nonlinear phase.

Another interesting feature of the pair/synchrotron instability is that the photon spectrum breaks at energies between 50 and 500 keV and this agrees generally well with the recent OSSE observations of Seyferts (see Maisack et al. 1993). While there are certainly other models to explain this feature (see Zdziarski et al. 1993, Titarchuk & Mastichiadis 1994), the fact that, on the one hand, we can explain such a feature in a simple way within the framework of our model and, on the other, reproduce the correct 2-10 keV spectral slope, makes the present approach promising.

A limitation of the method is that it is not capable of treating the details of radiation transfer through the source, using instead an escape probability formalism. This problem is common also to other models treating pair plasmas in the kinetic equation approach. Furthermore, the microphysical processes were treated using simple approximations and some processes such as Coulomb collisions and bremsstrahlung were neglected altogether. However, these processes are important only in a rather limited region of parameter space (see Svensson 1982 and Coppi 1992), which we avoid.

Other effects we do not try to include in this work include the possible radiation from protons escaping downstream and the effects such radiation might have on the acceleration zone. This is justified if, for example, escaping particles are accreted into the black hole. Finally, we do not attempt to treat the effects of neutrons produced in hadronic collisions and their possible feedback on the system. However, at least for the para-meter space where the pair/synchrotron instability dominates, protons lose the bulk of their energy in proton-photon pair production and neutrons are, therefore, unimportant for the evolution of the system.

*Acknowledgements.* We thank Peter Duffy, Demos Kazanas and Lev Titarchuk for stimulating discussions. A.M. thanks the Deutsche Forschungsgemeinschaft for support under Sonderforschungsbereich 328.

# References

Aharonian, F.A., Kirillov-Ugryumov, V.G., Kotov, Y.D., 1983, Astrophys. 19, 139

Akhiezer, A.I., Berestetskii, V.B., 1969, Quantum Electrodynamics, Nauka, Moscow

Atoyan, A.M., 1992, A&A 257, 465

Atoyan, A.M., 1992, A&A 257, 475

Axford, W.I., 1981, Proc. 17th. Int. Cosmic Ray Conf. (Paris) 12, 155

Begelman, M.C., de Kool, M., Sikora M., 1991, ApJ 382, 416

Begelman, M.C., Rudak, B., Sikora, M., 1990, ApJ 362, 38

Biermann, P. L., 1992, in Stenger, V.J. et al. (eds.), High Energy Neutrino Astrophysics, World Scientific, p. 86

Blumenthal, G.R., Gould, R.J., 1970, Rev. Mod. Phys. 42, 237

Blumenthal, G.R., 1971, Phys. Rev. 3D, 2308

Coppi, P.S., 1992, MNRAS 258, 657

Coppi, P.S., Blandford, R.D., 1990, MNRAS 245, 453

Cowsik, R., Lee, M.A., 1982, Proc. Roy. Soc. A 383, 409

Dermer, C.D., 1986, ApJ 307, 47

Done, C., Fabian, A.C., 1989, MNRAS 240, 81

Fabian, A.C., Blandford, R.D., Guilbert, P.W., Phinney, E.S., Cuellar, L., 1986, MNRAS 221, 931

Felten, J.E., Morrison, P., 1966, ApJ 146, 686

Ghisellini, G., Guilbert, P.W., Svennson, R., 1988, ApJLet 334, L5

Ginzburg, V.L., Syrovatskii, S.I., 1965, Ann. Rev. Astron. Astrophys. 3, 297

Giovanoni, P.M., Kazanas, D., 1990, Nat 345, 319

Guilbert, P.W., Fabian, A.W., Rees, M., 1983, MNRAS 205, 593

Hoyle, F., 1960, MNRAS 120, 338

Johnson, P.A., Mastichiadis, A., Protheroe, R.J., Stanev, T.S., Szabo, A.P., 1994, to appear in J. Phys. G

Jones, F.C., 1968, Phys. Rev. 167, 1159

Kardashev, N.S., 1962, Sov. Astron. J. 6, 317

Kazanas, D., Ellison, D.C., 1986, ApJ 304, 178

Kirk, J.G., Mastichiadis, A., 1989, A&A 213, 75

Kirk, J.G., Mastichiadis, A., 1992, Nat 360, 135 (KM92)

Kirk, J.G., Melrose, D.B., Priest, E., 1994, 'Plasma Astrophysics' Springer-Verlag, Berlin

Lightman, A.P., Zdziarski, A.A., 1987 ApJ 253, 842

Maisack, M., Johnson, W.N., Kinzer, R.L. et al., 1993, ApJLet 407, L61

Mannheim, K., Biermann, P.L., 1989, A&A 221, 211

Mastichiadis, A., Kirk, J.G., 1992, in Stenger, V.J. et al. (eds), High Energy Neutrino Astrophysics, World Scientific, p. 63

Mastichiadis, A., Protheroe, R.J., 1990, MNRAS 246, 279

McCray, R., 1969, ApJ 156, 329

Payne, D.G., Blandford, R.D., 1981, MNRAS 196, 781

Pounds, K.A., Nandra, K., Stewart, G.C., George, I.M., Fabian, A.C., 1989, Nat 344, 132

Protheroe, R.J., Kazanas, D., 1983 ApJ 265, 620

Protheroe, R.J., Szabo, A.P., 1992, Phys. Rev. Lett. 69, 2885

Rees, M.J., 1984, Ann. Rev. Astron. Astrophys. 22, 471

Rothschild, R.E., et al., 1983, ApJ 269, 423

Rybicki, G.R., Lightman, A.P., 1979, Radiative Processes in Astrophysics, McGraw-Hill, New York

Schlickeiser, R., 1984, A&A 136, 227

Schneider, P., Bogdan, T.J., 1989, ApJ 347, 496

Sikora, M., Begelman, M.C., 1992, in Stenger, V.J. et al. (eds), High Energy Neutrino Astrophysics, World Scientific, p. 114

Sikora, M., Begelman, M.C., Rudak, B., 1989, ApJLet 341, L33

Sikora, M., Kirk, J.G., Begelman, M.C., Schneider, P., 1987, ApJLet 320, L81

Stecker, F.W., 1968, Cosmic Gamma Rays, NASA SP-249

Stecker, F.W., Done, C., Salamon, M.H., Sommers, P., 1991, Phys. Rev. Lett. 66, 2697

Stern, B.E., Svensson, R., 1991, in Zdziarski, A.A., Sikora, M. (eds), Relativistic Hadrons in Cosmic Compact Objects, Springer-Verlag,

Stern, B.E., Svensson, R., Sikora, M., 1991, in Miller, H.R., Wiita, P.J. (eds), Variability of Active Galactic Nuclei, Cambridge University Press, p. 229

Stern, B.E., Sikora, M., Svensson, R., 1992, in Holt, S.S., Neff, S.G. (eds), Testing the AGN paradigm, American Institute of Physics, New York, p.313

Svensson, R., 1982, ApJ 258, 321

Svensson, R., 1987, MNRAS 227, 403

Svensson, R., 1989, in Brinkmann, W. (ed), Physical Processes in Hot Cosmic Plasmas, Kluwer Academic, p. 357

Szabo, A.P., Protheroe, R.J., 1992, in Stenger, V.J. et al. (eds), High Energy Neutrino Astrophysics, World Scientific, p. 24

Szabo, A.P., Protheroe, R.J., 1994, to appear in Astroparticle Phys

Titarchuk, L., Mastichiadis, A. 1994, ApJLet, in press

Turner, T.J., Pounds, K.A., 1989, MNRAS 240, 833

Webb, G.M., Bodgan, T.J., 1987, ApJ 320, 683

Zdziarski, A.A., 1989, ApJ 342, 1108

Zdziarski, A.A., Ghisellini, G., George, I.M., Svennson, R., Fabian, A.C., Done, C., 1990, ApJLet 363, L1

Zdziarski, A.A., Lightman, A.P., Maciolec-Niedzwieski, A., 1993, ApJLet 414, L93